# DISCUSSION OF "2004 IMS MEDALLION LECTURE: LOCAL RADEMACHER COMPLEXITIES AND ORACLE INEQUALITIES IN RISK MINIMIZATION" BY V. KOLTCHINSKII

By Sara van de Geer

*ETH Zürich*

**1. Introduction.** This paper unifies and extends important theoretical results on empirical risk minimization and model selection. It makes extensive and efficient use of new probability inequalities for the amount of concentration of the (possibly symmetrized) empirical process around its mean. The results are very subtle and very pleasing indeed, as they show that oracle inequalities exist for very general problems.

There are in my view two aspects which need special attention. First, the paper assumes that the loss functions $f \in \mathcal{F}$ satisfy $|f| \leq K$ for some fixed constant $K$. Let us call this the *uniform bound condition* (Condition B below). Second, it is not clear how the approach used will work in practice: the estimators depend on (unspecified) constants which may be too large for all practical purposes, and moreover, it is difficult to explain the method to nonspecialists. This discussion will address these two problems.

We reformulate some of the results as starting point for possible extensions or alternative approaches. For transparency, we will invoke simple, and not the most general, assumptions.

Section 2 in this discussion presents a distribution-dependent upper bound for the excess risk, replacing the uniform bound condition by convexity conditions and a bound on the renormalized loss functions (Condition BB).

The background of Section 3 in this discussion is the question whether cross-validation can be a more user-friendly model selection method than applying bounds in terms of Rademacher complexities. We first study why (data-dependent) upper and lower bounds for excess risks are useful when aiming at oracle behavior in model selection. We then show that when the margin behavior of the excess risk in each model is known, cross-validation can lead to oracle behavior.

Let us now first introduce our notation, following mostly that of the paper. Assume the observations $X_1, \ldots, X_n$ are i.i.d. copies of a random variable $X \in S$ with distribution $P$. Let $\mathcal{F}$ be a given class of functions $f$ on $S$. The empirical risk minimizer is

$$\hat{f} := \arg\min_{f \in \mathcal{F}} P_n f,$$

and its theoretical counterpart is

$$\bar{f} := \arg\min_{f \in \mathcal{F}} P f.$$

We assume for simplicity that the minimizers exist. The excess risk at $f$ is defined as $\mathcal{E}(f) := Pf - P\bar{f}$.

A distribution-dependent upper bound for $\mathcal{E}(\hat{f})$ depends on two ingredients, which we refer to as (1) the empirical process behavior and (2) the margin behavior.

Let

$$\sigma^2(f) := Pf^2 - (Pf)^2,$$

and let

$$\mathcal{F}_\sigma := \{f \in \mathcal{F} : \sigma(f - \bar{f}) \leq \sigma, |f - \bar{f}| \leq 1\}.$$

Consider the maximal increment of the empirical process

$$Z(\sigma) := \sup_{f \in \mathcal{F}_\sigma} |P_n(f - \bar{f}) - P(f - \bar{f})|.$$

The empirical process behavior is the behavior of $\mathbf{E}Z(\sigma)$ as function of $\sigma$. Bousquet's inequality [1] implies that for all $\bar{\varepsilon} > 0$,

$$(1) \quad \mathbf{P}\left(Z(\sigma) \geq (1+\bar{\varepsilon})\mathbf{E}Z(\sigma) + \sigma\sqrt{\frac{2t}{n}} + \left(\frac{1}{3} + \frac{1}{\bar{\varepsilon}}\right)\frac{t}{n}\right) \leq e^{-t} \qquad \forall t > 0.$$

The margin behavior of $Pf$ is the behavior of $\mathcal{E}(f)$ for $\sigma(f - \bar{f})$ small. This is described by

$$D(\delta) = \sup\{\sigma(f - \bar{f}) : f \in \mathcal{F} : \mathcal{E}(f) \leq \delta\}.$$

Condition A below (and also Conditions CC, C and $\{C(k)\}$) imposes certain conditions on the margin behavior.

We now combine empirical process behavior and margin behavior in the quantity $W_t(D(\delta))$, where

$$W_t(\sigma) = \frac{8}{5}\mathbf{E}Z(\sigma) + \sigma\sqrt{\frac{2t}{n}}$$

is inspired by (1). We set (quite arbitrarily) the value of $\bar{\varepsilon}$ at $\bar{\varepsilon} = 3/5$, that is, we do not attempt here to optimize our constants (for simplicity).

Lemma 1 below (and its proof) is a slight variant of the approach used in the paper. It will be applied in Lemmas 2 and 3 to obtain distribution-dependent bounds. In the lemma we invoke *conjugates*. The conjugate of a convex nondecreasing function $G$ on $[0, \infty)$ with $G(0) = 0$ is defined as the function $H(v) = \sup_{u \geq 0}[uv - G(u)]$.

Let us now fix some $t > 0$, and assume

CONDITION A. *There exists a strictly increasing concave upper bound $\psi_t(\delta)$ of $W_t(D(\delta))$, satisfying*

(i) $\psi_t^{-1}$ *has conjugate* $H_t$,
(ii) $\psi_t(\delta)/\delta$ *is nonincreasing in* $\delta$.

The conjugate $H_t(z)$ corresponds roughly speaking to a bound for the $\sharp$-transform $U_{n,t}^\sharp(\frac{1}{z})$ defined in the paper. We use conjugates to clarify the relation with our margin Conditions CC, C and $\{C(k)\}$.

For $\delta > 0$, we let

$$\mathcal{F}_1^\delta = \{f \in \mathcal{F} : |f - \bar{f}| \leq 1, \mathcal{E}(f) > \delta\}.$$

LEMMA 1. *Suppose Condition A. Then we have for all $q > 1$, $\varepsilon > 0$, $t > 0$ and $\delta > 0$,*

$$\mathbf{P}\left(\sup_{\mathcal{F}_1^\delta}\left|\frac{(P_n - P)(f - \bar{f})}{\varepsilon(H_t(\frac{1}{\varepsilon}) + \mathcal{E}(f)) + \frac{2t}{qn}}\right| \geq q\right) \leq \log_q \frac{q}{\delta} e^{-t}.$$

PROOF. Define $\delta_j = q^{-j}$, $j = 0, 1, \ldots$. Then for any $\delta > 0$, and for $\delta_J \geq \delta$,

$$\mathbf{P}\left(\sup_{\mathcal{F}_1^\delta}\left|\frac{(P_n - P)(f - \bar{f})}{\psi_t(\mathcal{E}(f)) + \frac{2t}{qn}}\right| \geq q\right) \leq \sum_{j=0}^J \mathbf{P}\left(Z(D(\delta_j)) \geq q\psi_t(\delta_{j+1}) + \frac{2t}{n}\right)$$

$$\leq \sum_{j=0}^J \mathbf{P}\left(Z(D(\delta_j)) \geq \psi_t(\delta_j) + \frac{2t}{n}\right) \leq \log_q \frac{q}{\delta} e^{-t}.$$

The result now follows, since for any $\varepsilon > 0$ and any $x > 0$, $\psi_t(x) \leq \varepsilon(H_t(\frac{1}{\varepsilon}) + x)$. □

The following lemma presents an upper bound for the excess risk. The lemma (and its proof) is a simplified (less general) version of Theorem 1 (and its proof) in the paper. We present the lemma here in order to allow comparison with the extension to the case where a uniform bound on the functions in $\mathcal{F}$ is not available (see Lemma 3 below).

Assuming a uniform bound condition, a distribution-dependent bound for the excess risk takes the form

$$\delta_{t,n} := \frac{q\varepsilon}{1-q\varepsilon} H_t\left(\frac{1}{\varepsilon}\right) + \frac{2t}{(1-q\varepsilon)n}. \tag{2}$$

CONDITION B. *We have $|f - \bar{f}| \leq 1$ for all $f \in \mathcal{F}$.*

LEMMA 2. *Suppose Conditions A and B. Then for all $q > 1$, $0 < \varepsilon < 1/q$ and $\delta \geq \delta_{t,n}$,*

$$\mathbf{P}(\mathcal{E}(\hat{f}) > \delta) \leq \log_q \frac{q}{\delta} e^{-t}.$$

PROOF. Let $\delta > 0$ and let $E$ be the event

$$\sup_{f \in \mathcal{F} : \mathcal{E}(f) > \delta} \left| \frac{(P_n - P)(f - \bar{f})}{\varepsilon(H_t(\frac{1}{\varepsilon}) + \mathcal{E}(f)) + \frac{2t}{qn}} \right| \leq q.$$

Since

$$\mathcal{E}(\hat{f}) \leq |(P_n - P)(\hat{f} - \bar{f})|,$$

we know that on $E$,

$$\mathcal{E}(\hat{f}) \leq \delta_{t,n} \wedge \delta.$$

Therefore, when $\delta \geq \delta_{t,n}$

$$\mathbf{P}(\mathcal{E}(\hat{f}) > \delta) \leq 1 - \mathbf{P}(E) \leq \log_q \frac{q}{\delta} e^{-t}. \qquad \square$$

**2. The case of possibly unbounded functions.** In this section, we assume that $\mathcal{F}$ is indexed by a parameter $\theta$ in some space $\Theta$:

$$\mathcal{F} = \{\gamma_\theta : \theta \in \Theta\}.$$

We moreover assume that $\Theta$ is a convex subset of a normed vector space with norm $\tau$, and that $\theta \mapsto \gamma_\theta(x)$ is convex for all $x \in S$. We let $\bar{f} = \gamma_{\bar{\theta}}$ and $\hat{f} = \gamma_{\hat{\theta}}$.

CONDITION BB. *Suppose that for some $\eta_n > 0$,*

$$\eta_n |\gamma_\theta - \gamma_{\bar{\theta}}| \leq \tau(\theta - \bar{\theta}) \vee \eta_n.$$

We also need a margin condition. Let $\Theta_1 := \{\theta \in \Theta : |\gamma_\theta - \gamma_{\bar{\theta}}| \leq 1\}$.

CONDITION CC. *For some increasing function $\mathbf{D}$,*

$$\mathbf{D}(\mathcal{E}(\gamma_\theta)) \geq \tau(\theta - \bar{\theta}) \qquad \forall \theta \in \Theta_1.$$

Define now
$$\tau_n := \mathbf{D}(\delta_{t,n}),$$
where $\delta_{t,n}$ is given in (2).

LEMMA 3. *Let $q > 1$ and $0 < \varepsilon < 1/q$ be arbitrary. Assume Conditions A, BB and CC, and that $\tau_n \leq \eta_n/2$. Then we have*
$$\mathbf{E}(\mathcal{E}(\hat{f}) > \delta_{t,n}) \leq \log_q \frac{q}{\delta_{t,n}} e^{-t}.$$

PROOF. Define
$$\tilde{\theta} = \alpha\hat{\theta} + (1-\alpha)\bar{\theta},$$
with
$$\alpha = \frac{2\tau_n}{2\tau_n + \tau(\hat{\theta} - \bar{\theta})}.$$
Then
$$|\gamma_{\tilde{\theta}} - \gamma_{\bar{\theta}}| = \frac{2\tau_n|\gamma_{\hat{\theta}} - \gamma_{\bar{\theta}}|}{2\tau_n + \tau(\hat{\theta} - \bar{\theta})} \leq \frac{2\tau_n|\gamma_{\hat{\theta}} - \gamma_{\bar{\theta}}|}{\tau(\hat{\theta} - \bar{\theta})} \leq 1.$$
Moreover, by the convexity of $P_n(\gamma_\theta)$, for $\tilde{f} := f_{\tilde{\theta}}$,
$$P_n(\tilde{f}) \leq \alpha P_n(\hat{f}) + (1-\alpha)P_n(\bar{f}) \leq P_n(\bar{f}).$$
This implies
$$\mathcal{E}(\tilde{f}) \leq |(P_n - P)(\tilde{f} - \bar{f})|.$$
Let $E_n$ be the event
$$\sup_{\mathcal{F}_1^{\delta_{t,n}}} \left| \frac{(P_n - P)(f - \bar{f})}{\varepsilon(H_t(\frac{1}{\varepsilon}) + \mathcal{E}(f)) + \frac{2t}{qn}} \right| \leq q.$$
By the same arguments as in Lemma 2, we have on $E_n$
$$\mathcal{E}(\tilde{f}) \leq \delta_{t,n}.$$
But then
$$\tau(\tilde{\theta} - \bar{\theta}) \leq \mathbf{D}(\delta_{t,n}) = \tau_n.$$
Hence
$$\tau(\hat{\theta} - \bar{\theta}) \leq 2\tau_n.$$
But then also
$$|\gamma_{\hat{\theta}} - \gamma_{\bar{\theta}}| \leq 1.$$
So on $E_n$ also
$$\mathcal{E}(\hat{f}) \leq \delta_{t,n}. \qquad \square$$

**3. Model selection.** Consider now a set of models $\{\mathcal{F}_k\}$, with $\mathcal{F}_k \subset \mathcal{F}$ for all $k$. Let

$$f_* := \arg\min_{f \in \mathcal{F}} Pf, \qquad \bar{f}_k := \arg\min_{f \in \mathcal{F}_k} Pf,$$

and denote the empirical risk minimizer in model $k$ by

$$\hat{f}_k := \arg\min_{f \in \mathcal{F}_k} P_n f.$$

We moreover define the excess risk at $\hat{f}_k$ within the model $k$ as

$$\mathcal{E}_k := P(\hat{f}_k - \bar{f}_k)$$

and the "empirical" excess risk at $\bar{f}_k$,

$$\hat{\mathcal{E}}_k := P_n(\bar{f}_k - \hat{f}_k).$$

The overall excess risk at $f$ is

$$\mathcal{E}_*(f) := P(f - f_*).$$

Let $\hat{\pi}(k)$ be some (data-dependent) penalty, and

$$\hat{k} := \arg\min\{P_n \hat{f}_k + \hat{\pi}(k)\},$$

assuming for simplicity that the minimum exists. It is important to find good estimates of the ("empirical") excess risks, because we can use these in the construction of a penalty $\hat{\pi}$. To clarify why, we reformulate Lemma 4 in the paper, combined with part of the proof of its Theorem 6. We also impose its margin condition (5.3), which we refer to as Condition C.

CONDITION C. *We have*

$$\mathcal{E}_*(f) \geq \phi[\sigma(f - f_*)] \qquad \forall f \in \mathcal{F},$$

*where $\phi$ is a function with conjugate $\phi^*$.*

LEMMA 4. *Assume Conditions B and C. Let $0 < \varepsilon < 1$ be arbitrary and let $\{t_k\}$ be an arbitrary positive sequence. Define for all $k$,*

$$\alpha(k) := \varepsilon \phi^*\left(\sqrt{\frac{t_k}{n\varepsilon^2}}\right) + \frac{t_k}{n}.$$

*Then*

$$\mathbf{P}\left(\mathcal{E}_*(\hat{f}_{\hat{k}}) > \frac{1}{(1-\varepsilon)^2} \min_k \{\mathcal{E}_*(\bar{f}_k) + (1-\varepsilon)[\alpha(k) + \hat{\pi}(k)]\}\right)$$
$$\leq \sum_k e^{-t_k} + \mathbf{P}(\exists k : \hat{\pi}(k) < \hat{\mathcal{E}}_k + (1-\varepsilon)\mathcal{E}_k + \alpha(k)).$$

PROOF. By Bernstein's inequality, with probability at least $1 - e^{-t_k}$,

$$|(P_n - P)(\bar{f}_k - f_*)| \leq \sqrt{\frac{2t_k}{n}}\sigma(\bar{f}_k - f_*) + \frac{t_k}{n}$$
$$\leq \varepsilon\phi[\sigma(\bar{f}_k - f_*)] + \alpha(k) \leq \varepsilon\mathcal{E}_*(\bar{f}_k) + \alpha(k).$$

But then

$$(1-\varepsilon)\mathcal{E}_*(\bar{f}_k) \leq P_n(\bar{f}_k - f_*) + \alpha(k)$$

and

$$P_n(\bar{f}_k - f_*) \leq (1+\varepsilon)\mathcal{E}_*(\bar{f}_k) + \alpha(k) \leq \frac{1}{1-\varepsilon}\{\mathcal{E}_*(\bar{f}_k) + (1-\varepsilon)\alpha(k)\}.$$

Let $E$ be the set where it holds for all $k$ that

$$\mathcal{E}_*(\bar{f}_k) \leq \frac{1}{1-\varepsilon}\{P_n(\bar{f}_k - f_*) + \alpha(k)\}, P_n(\bar{f}_k - f_*)$$
$$\leq \frac{1}{1-\varepsilon}\{\mathcal{E}_*(\bar{f}_k) + (1-\varepsilon)\alpha(k)\}$$

and

$$\hat{\pi}(k) \geq \hat{\mathcal{E}}_k + (1-\varepsilon)\mathcal{E}_k + \alpha(k).$$

We have on $E$,

$$P_n(\hat{f}_{\hat{k}} - f_*) + \hat{\pi}(\hat{k}) = \min_k\{P_n(\hat{f}_k - f_*) + \hat{\pi}(k)\} \leq \min_k\{P_n(\bar{f}_k - f_*) + \hat{\pi}(k)\}$$
$$\leq \frac{1}{1-\varepsilon}\min_k\{\mathcal{E}_*(\bar{f}_k) + (1-\varepsilon)[\alpha(k) + \hat{\pi}(k)]\}.$$

We also have on $E$,

$$\mathcal{E}_*(\bar{f}_{\hat{k}}) \leq \frac{1}{1-\varepsilon}\{P_n(\bar{f}_{\hat{k}} - f_*) + \alpha(\hat{k})\} = \frac{1}{1-\varepsilon}\{P_n(\hat{f}_{\hat{k}} - f_*) + \hat{\mathcal{E}}_{\hat{k}} + \alpha(\hat{k})\}.$$

Hence, on $E$,

$$\mathcal{E}_*(\hat{f}_{\hat{k}}) = \mathcal{E}_{\hat{k}} + \mathcal{E}_*(\bar{f}_{\hat{k}}) \leq \frac{1}{1-\varepsilon}\{P_n(\hat{f}_{\hat{k}} - f_*) + \hat{\mathcal{E}}_{\hat{k}} + \alpha(\hat{k}) + (1-\varepsilon)\mathcal{E}_{\hat{k}}\}$$
$$\leq \frac{1}{1-\varepsilon}\{P_n(\hat{f}_{\hat{k}} - f_*) + \hat{\pi}(\hat{k})\}$$
$$\leq \frac{1}{(1-\varepsilon)^2}\min_k\{\mathcal{E}_*(\bar{f}_k) + (1-\varepsilon)[\alpha(k) + \hat{\pi}(k)]\}. \quad \square$$

The above lemma indicates that one needs bounds for the ("empirical") excess risk, as well as knowledge of the margin behavior, that is, of the function $\phi$. This is also the message of the paper, and it is the reason why it studies such bounds. Theorem 2 in the paper handles the empirical excess risk $\hat{\mathcal{E}}_k$. Its Theorem 3 shows that one can estimate the distribution-dependent upper bounds for the excess risk $\mathcal{E}_k$. The latter is done using Rademacher complexities, which are based on symmetrized versions of the empirical process.

Symmetrization inequalities are based on comparing $P_n$ with an independent copy $P'_n = \frac{1}{n}\sum_{i=1}^n \delta_{X'_i}$, where $\{X'_n, \ldots, X'_n\}$ is a second sample independent of $\{X_1, \ldots, X_n\}$. The question arises whether simple data splitting can also be used to estimate $\mathcal{E}_k$ and $\hat{\mathcal{E}}_k$. Suppose indeed we have observed $\{X'_n, \ldots, X'_n\}$ in addition to $\{X_1, \ldots, X_n\}$. We let

$$\hat{f}'_k = \arg\min_{f \in \mathcal{F}_k} P'_n f.$$

Moreover, we define

$$\mathcal{E}'_k = P(\hat{f}'_k - \bar{f}_k), \qquad \hat{\mathcal{E}}'_k = P'_n(\bar{f}_k - \hat{f}'_k).$$

We now assume the following margin condition:

CONDITION $\{C(k)\}$.   For all $k$,

$$P(f - \bar{f}_k) \geq \phi_k[\sigma(f - \bar{f}_k)] \qquad \forall f \in \mathcal{F}_k,$$

where $\phi_k$ has conjugate $\phi_k^*$.

Define now for all $k$ the (truly) empirical quantities

$$\hat{\beta}(k) := (P'_n - P_n)(\hat{f}_k - \hat{f}'_k).$$

LEMMA 5.   *Assume Conditions* B *and* $\{C(k)\}$. *Let* $0 < \varepsilon < 1$ *be arbitrary and let* $\{t_k\}$ *be an arbitrary positive sequence. Define*

$$\gamma(k) = \varepsilon \phi_k^*\left(\sqrt{\frac{2t_k}{n\varepsilon^2}}\right) + \frac{t_k}{n}.$$

*Then with probability at least* $1 - \sum_k e^{-t_k}$, *we have for all* $k$

$$\hat{\beta}(k) + 2\gamma(k) \geq (1-\varepsilon)\{\mathcal{E}'_k + \mathcal{E}_k\} + \hat{\mathcal{E}}'_k + \hat{\mathcal{E}}_k.$$

PROOF.   By Bernstein's inequality, conditionally on $X_1, \ldots, X_n$, we have with probability at least $1 - \frac{1}{2}e^{-t_k}$, that

$$(P - P'_n)(\hat{f}_k - \bar{f}_k) \leq \sqrt{\frac{2t_k}{n}}\sigma(\hat{f}_k - \bar{f}_k) + \frac{t_k}{n}.$$

But then,

$$(P - P'_n)(\hat{f}_k - \bar{f}_k) \leq \varepsilon \mathcal{E}_k + \gamma(k) \quad \text{or} \quad P'_n(\hat{f}_k - \bar{f}_k) \geq (1-\varepsilon)\mathcal{E}_k - \gamma(k).$$

Similarly for $P_n(\hat{f}'_k - \bar{f}_k)$.

Let $E$ be the set where it holds that for all $k$

$$P'_n(\hat{f}_k - \bar{f}_k) \geq (1-\varepsilon)\mathcal{E}_k - \gamma(k)$$

and

$$P_n(\hat{f}'_k - \bar{f}_k) \geq (1-\varepsilon)\mathcal{E}'_k - \gamma(k).$$

Then on $E$, we have

$$\begin{aligned}(P'_n - P_n)(\hat{f}_k - \hat{f}'_k) &= (P'_n - P_n)(\hat{f}_k - \bar{f}_k) + (P_n - P'_n)(\hat{f}'_k - \bar{f}_k) \\ &= P'_n(\hat{f}_k - \bar{f}_k) + \hat{\mathcal{E}}_k + P_n(\hat{f}'_k - \bar{f}_k) + \hat{\mathcal{E}}'_k \\ &\geq (1-\varepsilon)\{\mathcal{E}_k + \mathcal{E}'_k\} + \hat{\mathcal{E}}_k + \hat{\mathcal{E}}'_k - 2\gamma(k). \quad \square\end{aligned}$$

It follows that if the margin behavior of all models $\{\mathcal{F}_k\}$ and $\mathcal{F}$ is known, one may take as data-dependent penalty

(3) $$\hat{\pi}(k) = \hat{\beta}(k) + \alpha(k) + 2\gamma(k).$$

One can then apply Lemma 4. One may proceed by proving a distribution-dependent upper bound for this choice of $\hat{\pi}(k)$ (this bound actually follows from the paper). The penalty clearly has as advantage that it is simple to implement. But as it requires the margin behavior to be known, there are many problems (e.g., classification) where it cannot be used. In the paper, Theorems 5 and 6, only the margin behavior of the model $\mathcal{F}$ is assumed to be known. Thus, one might say that by estimating the upper bounds for the excess risks (using Rademacher complexities), instead of the excess risks themselves, one overcomes the problem of not knowing the margin behavior of *all* models $\mathcal{F}_k$.

The paper moreover shows that by replacing the penalization method by a comparison method, the margin problem can be solved. Another promising approach is in my view the use of $l_1$-type penalties, but these are only defined within the context of linear parameter spaces.

SEMINAR FÜR STATISTIK
ETH ZENTRUM
LEO D11
8092 ZÜRICH
SWITZERLAND
E-MAIL: geer@stat.math.ethz.ch